\documentclass[twocolumn,5p]{elsarticle}
\usepackage{amsmath,amsfonts,epsfig}

\newcommand{\Z}{{\mathbb Z}}

\begin{document}

\title{Towards the MSSM from F-theory}
\author{Kang-Sin Choi}
\ead{kschoi@gauge.scphys.kyoto-u.ac.jp}
\author{Tatsuo Kobayashi}
\ead{kobayash@gauge.scphys.kyoto-u.ac.jp}
\address{Department of Physics, Kyoto University, Kyoto
606-8502, Japan}
%\preprint{KUNS-2252}

\begin{abstract}
We study the MSSM in F-theory. Its group is the commutant to a
structure group $SU(5) \times U(1)_Y$ of a gauge bundle in $E_8$.
The spectrum contains three generations of quarks and leptons
plus vectorlike electroweak and colored Higgses.
The minimal MSSM Yukawa couplings with
matter parity is obtained at the renormalizable level.
\end{abstract}

\maketitle

\newpage

\section{Introduction}

The purpose of this letter is to describe the Minimal Supersymmetric
Standard Model (MSSM) in F-theory. String theory provides 
self-consistent ultraviolet completion of the gauge theory. For
example, the constraint of anomaly freedom in the
low energy theory originates from the finiteness of string one-loop
amplitude.

F-theory is defined by identifying $S$-duality of IIB theory with
the symmetry of a torus, lifting the gauge symmetry to geometry
\cite{Vafa:1996xn}.
To have four dimensional theory with ${\cal N}=1$ supersymmetry, we
compactify F-theory on Calabi--Yau fourfold that is elliptic
fibered. Gauge bosons are localized on a complex surface $S$,
along which the fiber is singular.
The structure of the singularities has 
correspondence to that of the corresponding group,
so that the symmetry breaking and
enhancement are described by geometric transition \cite{KV}.
Analogous to the bifundamental representations at D-brane intersections,
matter fermions come from the `off-diagonal' components
under branching of the gaugino of $E_8$ \cite{Donagi:2009ra}, localized along
matter curves \cite{Morrison:1996pp}.

So far there have been constructions of Grand
Unified Theories (GUTs) mainly based on the simple group $SU(5)$, and
further attention has been paid on the flavor sector
\cite{Beasley:2008dc,Donagi:2008kj,Heckman:2010bq}.
Compared to field theoretic GUT, presumably its best merit is that we do not
need adjoint Higgses for breaking down to SM. Instead, we can
turn on a flux in the hypercharge $U(1)_Y$ direction, evading
complicated GUT vacuum configuration. This flux does not break
hypercharge interaction, if
a certain topological condition is satisfied \cite{Donagi:2008kj}.
However, further breaking down to SM can potentially lead to chiral
multiplet containing $X$-boson, the off-diagonal component of the adjoint of
$SU(5)$, by the above mechanism localizing light matters.

\begin{figure}[h]
\begin{center}
\includegraphics[height=1cm]{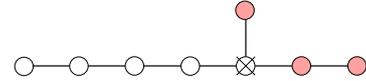}
\end{center}
\caption{The Standard Model group (filled) is obtained as the commutant to
  $SU(5)_\bot \times U(1)_Y$ background in $E_8$.}
\label{f:e3}
\end{figure}

Another obvious approach that we take here is to start with the group
$SU(3)\times SU(2) \times U(1)_Y$.
The SM group lies along the series of exceptional groups of $E_n$-type, and
we take $E_3 \times U(1)_Y$ group inside $E_8$, shown in
Fig. \ref{f:e3}, guaranteeing the correct field
contents and quantum numbers. Such $E_n$-series is naturally predicted
by F-theory and heterotic string.
For this we need a description on semisimple groups \cite{Choi:2009tp,KM}.
If $U(1)_Y$ is constructed via geometry and there is no flux along 
this part, embeddability to $SU(5)$ GUT singularity guarantees anomaly
free spectrum and we do not worry about its
breaking by Green--Schwarz mechanism.

\section{The Standard Model surface}

A gauge group is described by a singular fiber sharing the same name, which is read off from Tate's table \cite{Bershadsky:1996nh}.
We claim that the singularity describing the SM group is
\begin{equation} \label{weieq} \begin{split}
%y^2 = x^3 + b_5 xy + b_4 x^2 z + b_3 y z^2 + b_2 x z^3 + b_0 z^5.
y^2 & = x^3 + (b_5 + b_4 a_1) xy + (b_3 + b_2 a_1)(a_1 b_5+ z)zy  \\
& +(b_4 + b_3 a_1) x^2 z+ (b_2 - b_0 a_1^2)(a_1 b_5 +  z)x z^2  \\
 & + b_0 (a_1 b_5 +  z )^2 z^3.
  \end{split}
\end{equation}
For the total space to be Calabi--Yau, the equation should satisfy
topological conditions: $z,a_1,b_n$ are respectively
sections of ${\cal O}(S),{\cal O}(K_B+S),{\cal O}((n-6)K_B+(n-5)S)$ in
%We mean $K_B,N_{S/B}$ be respectively the canonical bundle of 
the base $B$ of elliptic fibration, and $K_B$ is its canonical bundle.
The surface $S$ is located at $z=0$, since the vanishing discriminant of (\ref{weieq})
\begin{equation} \label{SMdisc}
\Delta =(b_5+a_1b_4)^3 a_1^2 b_5^2 P_{d_\circ^c} P_{u_\circ^c} z^3  +  a_1 b_5 Q
 z^4 + O (z^5)
 \end{equation}
signals a singular fiber. We need globally defined sections in $B$, since the simple group components would be laid away from $S$.

Eq. (\ref{weieq}) is a deformation of $E_8$ singularity.
Surveying the degrees of the coefficients in (\ref{weieq}), Tate's
table shows it is 
generically an $SU(3)$ singularity. However the parameters are specially
tuned, so the actual symmetry is larger \cite{Choi:2009tp}. Vanishing
of each coefficient factor triggers gauge symmetry enhancement,
implying matter localization \cite{KV}. We see shortly that the
parameters $a_1 \equiv P_X$, $b_5 \equiv P_{q_\circ}$, $P_{d_\circ^c}$ and $P_{u_\circ^c}$
are respectively related to $X$-boson and subscripted quarks, which
are all the charged fields under $SU(3)$.  We can see, neither
$Q$ nor $O(z^5)$ is proportional to $a_1$
or $b_5$, hence $P_{d_\circ^c} P_{u_\circ^c}$ and $a_1 b_5 \to 0$ respectively
enhance the symmetry to $SU(4)$ and $SU(5)$. 

To see it contains also $SU(2)$, we change the variable $z' = z + a_1 b_5$, in which (\ref{SMdisc}) becomes
\begin{equation} \label{su2disc} \begin{split}
 \Delta = & \left( (b_5-a_1 b_4)^2 - 4 a_5^2 b_3 b_5 \right)^2  a_1^3
 b_5^3 P_{l_\circ} z^{\prime 2}  \\ 
 &+  a_1^2 b_5^2  P' z^{\prime 3} +  a_1 b_5 Q' z^{\prime 4} + O(z^{\prime 5}).
\end{split}\end{equation}
Here $P_{l_\circ}$ is related to the lepton doublet. Since neither of
$P'$,$Q'$ nor $O(z^{\prime 5})$ is proportional to $a_1$ or $b_5$, 
vanishing $P_{l_\circ}$ and $a_1 b_5$ respectively enhances the gauge symmetry
respectively to $SU(3)$ and $SU(5)$. The position $z'=0$ is off from the position of $S$ , as a `back-reaction' under the symmetry breaking from $SU(5)$, but it is still linearly equivalent to $S$. We can see, to make the embracing $SU(5)$ truly traceless, we need also a backreaction $z \to z-2 a_1 b_5/5$ to make the center of mass of the total brane lie on $S$.

The factorization structure of $P_{u_\circ^c} P_{d_\circ^c}$ and $P_X
P_{q_\circ} = a_1 b_5$ hints the existence of an additional $U(1)_Y$, otherwise we have only single factors for colored singlet and colored doublet,
respectively. However to guarantee the existence of such $U(1)_Y$ we should check the global factorization structure of (\ref{weieq}) \cite{Hayashi:2010zp}.
Above all, the most general deformation is by $a_1 \to 0$, implying the embedding of our symmetry
to a general $SU(5)$ GUT singularity in the literature \cite{Donagi:2009ra}.
Being its deformation,
the factorization structure of $SU(3) \times
SU(2) \times U(1)_Y$ is stably preserved against higher order
perturbation in $O(z^6)$.
Lying along the $E_n$ unification series $E_3 \times U(1)_Y \to E_4 \to E_8$, this SM group is the only
possibility.

\section{Matter contents}

The broken symmetry is
$SU(5)_\bot \times U(1)_Y$ `structure group'  whose commutant in $E_8$ is the 
Standard Model group \cite{BMW}. The spectral cover geometrically
describes it (or the dual to gauge bundle) satisfying supersymmetry conditions
\cite{Friedman:1997yq}. Roughly, it generalizes the notion of branes
whose symmetry broken by recombination, so sometimes called by flavor
brane stack, intersecting $S$ along the matter curves. 
It is described by factorized spectral cover $C = C_X \cup C_5:$
\begin{equation} \label{su5xu1cover}
  (a_0 s + a_1 )(b_0 s^5 - a_0^{-1} a_1 b_0 s^4   + b_2 s^3 + b_3 s^2 + b_4 s+ b_5)  = 0,
\end{equation}
where $s$ transforms as $-c_1 \equiv K_S$, the canonical bundle of $S$.
We also define $-t$ as the normal bundle to $S$ in $B$.
(\ref{su5xu1cover}) is embedded in a compact threefold \footnote{The compact (non-Calabi--Yau) threefold, 
a projectivized fiber $\pi:\mathbb{P}({\cal O}_S \oplus K_S) \to S$, with the trivial bundle ${\cal O}_S$ and 
the canonical bundle $K_S$.}, whose projection to $S$ we denote
$\pi$. 
The parameters are the ones used in (\ref{weieq}) projected on $S$: Using adjunction formula $b_m$ are sections of $(6-m)c_1 \equiv \eta - mc_1$ and  the combination $a_1 b_0/a_0$ is a section of $5 c_1-t$, with which the full equation for $C$ 
has no $s^5$ term showing the embeddability to a $SU(6)_\bot$
structure group \footnote{Since the trace part of $SU(5)_\bot$ is the
$U(1)_Y$ degree of freedom, it is sometimes called $S[U(5)\times
    U(1)]$.}. Using this embedding structure, one can be convinced
that the parameters in (\ref{weieq}) are the only possible combinations. 
We can recover the conventional $SU(5)$ GUT by $a_1 \to 0$ in
(\ref{SMdisc}), making its 
structure group traceless, reducing the above $C_5$ exactly
to the standard one used in $SU(5)$ GUT. 
The fields charged under two groups
satisfy the Green--Schwarz relations \cite{Sadov:1996zm,Choi:2009tp},
fixing $a_0=1$ to be trivial section.

We get the matter spectrum from the decomposition of the adjoint ${\bf
  248}$ of $E_8$
\begin{equation*}  \begin{split}
& {\bf (8,1,1)+(1,3,1)+(1,1,1)+(1,1,24)}\\
 & + X{\bf (3,2,1)}_{-5/6}   + q_\circ{\bf (3,2,5)}_{1/6}+d^c_\circ{\bf (\overline 3,1,10)}_{1/3}   \\
 &  +u^c_\circ{\bf (\overline 3,1,5)}_{-2/3} +
 l_\circ{\bf (1,2,10)}_{-1/2} +e^c_\circ{\bf (1,1,\overline{5})}_{-1},
  \end{split}
\end{equation*}
up to Hermitian conjugates.
%For identification
%\begin{equation*}  \begin{split}
%\bf (1,1,35) &\to {\bf (1,1,24)}+{\bf (1,1,1)} +e^c {\bf (1,1,5)}_1+ e {\bf (1,1,\overline 5)}_{-1} \\
%\bf (3,2,6) &\to q {\bf (3,2,5)}_{1/6} + X {\bf (3,2,1)}_{-5/6} \\
%\bf (\overline 3,1,15) &\to d^c {\bf (\overline 3,1,10)}_{1/3}+ u^c {\bf (\overline 3,1,5)}_{-2/3} \\
%\bf (1,2,20) &\to l {\bf (1,2,10)}_{-1/2} + \text{c.c.}.
%  \end{split}
%\end{equation*}
Matter fields are obtained as `off-diagonal' components of the
branching \cite{KV}.

They are further identified by local gauge symmetry enhancement directions.
For this, we parameterize their localizing curves using parameters
$t_1, t_2, \dots t_5$ having one-to-one correspondence with the five
weights of ${\bf 5}$ of $SU(5)_\bot$ and $t_6$ with $U(1)_Y$. 
For example, the $X$-boson appears from a local symmetry enhancement
to $SU(5)$, controlled by $ t_6 \to 0$, agreeing with the above. However the physical parameters are only the coefficients in the spectral cover (\ref{su5xu1cover}):
$b_n/b_0$ are the elementary symmetric polynomials of degree $k$, of
$t_1,t_2,t_3,t_4,t_5$, and $a_1 \equiv P_X \sim t_6$
\cite{Beasley:2008dc,Donagi:2009ra}.
$P_{q_\circ}$ is localized along the curve $b_5/b_0 \sim t_1 t_2 t_3 t_4
t_5 = 0$ and $P_{u_\circ^c} \sim \prod_{i=1}^5(t_i+t_6),
P_{d^c_\circ} \sim \prod_{i<j}^5 (t_i+t_j), P_{l_\circ} \sim \prod_{i<j}^5(t_i+t_j+t_6)$. 
The counting of $\bf 10$ and $\bf \overline{10}$ agrees
thanks to the relation
\begin{equation} \label{lepindex}
 t_i+t_j+t_6 =-t_k-t_l-t_m, \quad \epsilon_{ijklm} \ne 0.
\end{equation}
It is easy to see that the parameters $P_{m_\circ},m=X,q,u^c,d^c,l$, here calculated from
group theory, agree with those in (\ref{SMdisc}) and (\ref{su2disc}).
Again $t_6 \to 0$ reduces the matter curves to those of $SU(5)$, so the
$SU(5)$ GUT structure is preserved.

In the perturbative picture, parallel D-branes do
not intersect. Here, the base of elliptic fibration has a similar
structure generalizing Hirzeburch surface, where the zero section of
the fiber has nonzero
`self-intersection'. Thus $SU(3)$ and $SU(2)$ components intersect
yielding chiral fermions $X$ and $q$. 
In the sense of regarding the intersecting branes as connected
cycles \cite{Choi:2006hm}, this is not so different.

%\section{Yukawa couplings and GUT relations}

By dimensional reduction, we obtain Yukawa coupling from covariant
derivatives of gaugino of an enhanced group, as usual \cite{Beasley:2008dc}.
So the gauge invariance guarantees the presence of Yukawa couplings.
They are
\begin{equation} \begin{split}
l h_d e^c &:
(t_i+t_j+t_6)+(t_k+t_l+t_6)+(t_m-t_6) = 0, \\ \label{lhe}
  q h_u u^c &: (t_i)+(-t_i-t_j-t_6)+(t_j+t_6) = 0, \\
q h_d d^c &:
\left\{ \begin{matrix} (t_m)+(t_k+t_l+t_6)+(t_i+t_j) = 0,
   \\
(t_m)+(-t_m-t_i-t_j)+(t_i+t_j) = 0, \end{matrix} \right.
  \end{split}
\end{equation}
where all the indices run from 1 to 5, and all different.

%From the relations (\ref{qhu}), having repeated indices, the
%spectral covers for $q$, $h_u$ and $u^c$ should be related (see below).
Due to the relation (\ref{lepindex}), LHS's of (\ref{lhe})
vanish, and in each relation, we have two or more ways
of writing {\em the same} coupling. For example in the last line,
the first relation reduces the same relation to that of $SU(5)$ GUT,
in the limit $t_6 \to 0$. The second
relates the matter curves for all $q$, $h_d$, $d^c$, $h_u$ and
$u^c$. We may track this from the left-right symmetric models and
their extension (Pati--Salam and trinification) which relates up and
down sectors by $SU(2)_R$.
%If we include a singlet as the
%right-handed neutrino $\nu^c {\bf (1,1,1)}_0$,
%$
%\Sigma_{\nu^c}: t_i - t_j \to 0
%$
%we can consider the coupling
%\begin{equation} \label{lhnu}
%l h_u \nu^c: (t_i+t_j+t_6)+(-t_i-t_k-t_6)+(t_k-t_j), \ \epsilon_{ijk} \ne 0.
%\end{equation}
%As the quantum numbers and their transformation suggest, there are
%many singlet of moduli fields in the $U(5)_\bot$ instanton
%background.
We can switch between up-type and down-type fermions,
like doing between Georgi--Glashow $SU(5)$ and
flipped $SU(5)$.

\section{Matter curves and monodromy}

For a general spectral cover, we have monodromy condition
for connecting different matter curves \cite{Hayashi:2009ge}. In our
case, the $SU(5)_\bot$ has $S_5$ monodromy as reflected in the
symmetric polynomial relations of the coefficients:
The connected ones
form an orbit and are treated as the identical surface. It leads to just
one kind of lepton doublet, without distinguishing $l$, $h_d$ or
$h_u^c$.

To remedy this, we further mod out the others by $\Z_4$ monodromy
generated by the cyclic permutation of $(t_1,t_2,t_3,t_4
)$. It follows that, for
instance, $q$ is distinguished by different orbits
$\{t_1,t_2,t_3,t_4\}$ and $\{t_5\}$, whereas $X$
is still $\{t_6\}$.
There are distinct candidates for the matters. For example lepton or
Higgs belongs to one of the following $\Z_4$ orbit
\begin{equation*} 
l,h_d,h_u^c  : \{t_i + t_{i+1} + t_6 \}, \{t_1+t_3+t_6,t_2+t_4+t_6\},
\{ t_i+t_5+t_6 \},
\end{equation*}
with $i =1,\dots,4$.
To have lepton Yukawa coupling (\ref{lhe}), $h_d$ and $l$ must not share
$t_5$, therefore essentially we have two allowed cases for choosing
$SU(2)$ doublets. We choose the fields as in Table \ref{t:mcurve41}.
We named the colored exotics as colored Higgses $h_{c1}$ and $h_{c2}$.
Note also that, by $SU(5)$ unification structure the fields belonging to
a single multiplet has homologous curves.

We identify two Abelian symmetries generated by
\begin{equation} \begin{split}
 U(1)_Y: {\rm diag}(1,1,1,1,1,-5), \\
 U(1)_M: {\rm diag}(1,1,1,1,-4,0), \\
  \end{split}
\end{equation}
in the basis $\{t_1,t_2,t_3,t_4,t_5,t_6\}$.
The first is $U(1)_Y$ and the second is 
the famous `$B-L$' symmetry, commutant to $SU(5)$ inside
$SO(10)$, providing continuous version of matter parity.
Since both $h_u$ and $h_d^c$ have even $U(1)_M$ charges,
we can forbid lepton or baryon number
violating operators
\begin{equation} \label{rviolating}
lh_u,\ lle^c,\ lqd^c,\ u^c d^c d^c.
\end{equation}
We may also have operators
\begin{equation}  \label{xhd}
X h_u d^c,\ q' h_d^c h_{c1}^c,\ q' h_u h_{c2}.
\end{equation}
It turns out that neither $X$ nor $q'$ exists in four dimensions, for
later choice of the flux (see (\ref{zeromul}) below), so there will be
no problem.  

\begin{table}[t]
\begin{center}
\begin{tabular}{cccc} \hline
matter & matter curve &cycle on $S$ & $M$ \\ \hline
$X$    & $t_6 \to 0$         & $-c_1$ & $0$ \\
$q$    & $\prod t_i\to 0$          & $\eta-4c_1-x$ & $1$ \\
$q'$   & $t_5\to 0$          & $-c_1+x$  & $-4$ \\
$d^c $ & $\prod( t_i+t_5) \to 0$     & $ \eta - 4c_1+2x$ & $-3$ \\
$h_{c1}^c$   & $\prod(t_i+t_{i+2}) \to 0$ & $\eta - 2c_1-x$ & $2$ \\
$h_{c2}$ & $\prod(t_i+t_{i+1}) \to 0$ & $\eta - 4c_1-x$ & $2$ \\
$u^c $ & $\prod(t_i+t_6) \to 0$     & $\eta-4c_1-x$ & $1$ \\
$u' $  & $t_5+t_6 \to 0$     & $-c_1+x$ & $-4$ \\
$h_u^c$  & $\prod(t_i+t_{i+1}+t_6) \to 0$ & $\eta -4c_1-x$ & $2$ \\
$h_d$  & $\prod(t_i+t_{i+2}+ t_6) \to 0$ & $\eta -2c_1-x$ & $2$ \\
$l  $  & $\prod(t_i+t_5+t_6) \to 0$ & $\eta -4c_1+2x$ & $-3$ \\
$e^c $ & $\prod(t_i-t_6) \to 0$     & $\eta-4c_1-x$ & $1$ \\
$e' $  & $t_5-t_6 \to 0$     & $-c_1+x$ & $-4$ \\
\hline
\end{tabular}
\caption{Matter curves with $\Z_4$ monodromy.
All the indices take value in $\Z_4=\{1,2,3,4\}$ and are different.
The primed ones, charged exotics,
have always odd $M+2s$ charges.}
\label{t:mcurve41}
\end{center} \end{table}

The resulting spectral cover is further factorized $C_5 \to
C_{q'} \cup C_q$, thus (\ref{su5xu1cover}) becomes
\begin{equation} \label{factor411}
% \begin{split}
(s + a_1 )(d_0  s + d_1 )(e_0 s^4 + e_1 s^3  + (e_2'+e_2'') s^2  + e_3 s + e_4 )=0
% \end{split}
\end{equation}
with `traceless condition' $ a_1 d_0 e_0 + d_1 e_0
+ d_0 e_1 = 0$. 
The new covers are for $t_5 \sim d_1/d_0$, and $e_i$ will be again
related to elementary symmetry polynomials of degree $i$ out of
$t_1,t_2,t_3,t_4$. Only $e_2'/e_0 \sim (t_1+t_3)(t_2+t_4) $ and $e_2''/e_0
\sim t_1 t_3+t_2 t_4$ are nontrivially $\Z_4$ closed.
We have additional degree of freedom to choose $d_0$, a section $x \in
H_2(S,\Z)$. It follows $e_k$ are sections $e_k \sim \eta-kc_1-x$. 

As discussed around (\ref{lhe}), there are fields related by nonvanishing Yukawa couplings. They originate from the same spectral cover.
Since the defining conditions $t_i=0$ are stronger than $t_i+t_j=0$, 
%and $t_i+t_j+t_k=0$
we can derive all associated matter curves from the
`fundamental' spectral covers  $t_i=0$ for $C_X,C_{q'},C_q$
\cite{Donagi:2009ra,Marsano:2009gv}.
For example the matter
curve for $d^c$ is the common intersection between $C_q(t_i)$ and
$C_{q'}(-t_i)$, extracting the redundant component. 

{}From the symmetry enhancement directions, as shown in Table
\ref{t:mcurve41}, we can find the following classes of the matter curves:
\begin{equation} \label{curves} \begin{split}
   X&: C_X   \cap \sigma = -c_1 \cap \sigma \\
   q, u^c, e^c&: C_q   \cap \sigma = C_q \cap C_X = (\eta-4c_1-x) \cap \sigma \\
  q',u',e'&: C_{q'}\cap \sigma = C_{q'} \cap C_X=  (-c_1+x) \cap \sigma \\
 d^c, l&: (C_q-(\sigma+c_1)) \cap C_{q'} \\
   = & \textstyle \sigma \cap (\eta-4c_1+2x)+(\eta-c_1-x) \cap x \\
   h_{c1},h_d &:\textstyle C_q   \cap (2\sigma+c_1)\\
  =&\textstyle 2 \sigma \cap (\eta-2c_1-x)+(\eta-x) \cap c_1 \\
 h_{c2}^c,h_u^c &:\textstyle  (C_q -2\sigma)  \cap (C_q -4(\sigma+c_1)) \\
   = & \textstyle 2 \sigma \cap (\eta-4c_1-x)+(\eta-x) \cap (\eta-4c_1-x) 
  \end{split}
\end{equation}
where we omitted pullback. The associated matter curves have
$\sigma$-independent components that lie outside $S$ but on the
spectral cover. 
% We can collect (\ref{curves}) into unfactorized curves
%of (\ref{su5xu1cover})
%\begin{equation} \begin{split}
%& (C_5 - U) \cap (C_5 -3V) \\
%& = (C_q-2U) \cap (C_q-4V) + 2(C_q-V) \cap C_{q'} \\
%& \ + C_q \cap (U+V) + (C_{q'} - U)\cap (C_{q'}-V)
%\end{split}
%\end{equation}
%where the last term vanishes for $x=0$.
Since $a_0$ is trivial section, thus $C_X \sim \sigma$ and the matter curve with
and without $t_6$ are homologous, for $q$, $u^c$, $e^c$ for example.
Again this exhibits the unification relations to $SU(5)$ GUT.
More detailed calculation is to be found elsewhere \cite{CK}.

\section{Flux and spectrum}

To obtain four dimensional chiral spectrum, we should turn on a
magnetic flux $\gamma$ on the spectral cover $\gamma \in H^2(C,\Z)$
\cite{Friedman:1997yq,Abe:2009uz}.  With above factorization we turn on the
universal flux only on $C_q$ \footnote{There is no ramification on
  $C_{q'}$ and $C_X$ parts, so we can turn off fluxes on them, and the
  even (fourfold) cover of $C_q$ allows   
 that integral $\gamma_q$ corresponds to integral spectral line
 bundle \cite{Blumenhagen:2009yv}.}
\begin{equation}
 \gamma_{q}  = C_{q} \cap (4\sigma - \pi^*(\eta-4c_1)), \quad
 \gamma_{q'}=\gamma_{X}=0.
\end{equation}
We can show it has integral cohomology and traceless $p_{C *} \gamma =
0$ inside $SU(5)_\bot \times U(1)_Y$, where $p_C:C \to S$.
It follows that $q'$, $X$ and the associated matters are all neutral
to have zero multiplicity 
\begin{equation} \label{zeromul}
 n_X = n_{q'} = n_{u'} = n_{e'} =  0.
\end{equation}
We calculate the net number of fields, or the differences $n_f$
between the numbers of fermions $f$ and antifermions $f^c$ by
Riemann--Roch--Hirzebruch index theorem. It assumes the same form for
associated matter curves, namely the product of $4\sigma -
p_q^*(\eta-4c_1)$ with the curves in (\ref{curves}), letting $p_q$ the
projection $C_q \to S$ \cite{Hayashi:2008ba,Blumenhagen:2009yv}.  
For example, we obtain
\begin{equation*} \begin{split}
 n_q &=\sigma \cap  (\eta-4c_1-x) \cap(4\sigma -
p_q^*(\eta-4c_1))|_S\\
& = -\eta \cdot (\eta-4c_1 -x) \\
 n_{d^c}& =  (\sigma \cap (\eta-4c_1+2x)+  (\eta-c_1-x) \cap x) \\
  & \quad \cap  (4\sigma -p_q^*(\eta-4c_1))|_S \\
& = -\eta\cdot (\eta-4c_1+2x)+4(\eta-c_1-x)
\cdot x,
  \end{split}
\end{equation*}
where the dot product $\cdot$ means the intersection between the
divisors on $S$ and we used the Poincar\'e dual flux.
In fact, Kodaira vanishing theorem states that either chiral or
antichiral fermion has exclusively nonzero in most cases
\cite{Beasley:2008dc,Donagi:2008kj}.
%Note that $CPT$ conjugates are
%related by $t_i \leftrightarrow -t_i$ exchanging the orientation of
%the matter curve.
%\begin{equation} \label{6dgsrel} \begin{split}
% \sum_{R ,R'}  \ell(R){\rm dim}(R') \Sigma_{R,R'} |_S
%  &=\left \{ \begin{matrix}
%  9c_1 - 6t, R\in SU(3) \\
% 8c_1 - 6t, R \in SU(2)\end{matrix}\right. \\
% \sum_{R,R'} \ell(R)\ell(R') \Sigma_{R,R'}|_S & = -t,
%  \end{split}
%\end{equation}
%where $\ell(R)$ is the index of the representation $R$.
The anomaly cancellation condition for $SU(3)$, 
$2n_q  - n_{d^c} -n_{u^c}=
0$, requires also $ x=0 \  \text{ or }\ 4x=5\eta-c_1. $
We choose $x=0$ to have the spectrum
\begin{equation} \begin{split}
 & n_q = n_{d^c} = n_{u^c} = n_l = n_{e^c}=-\eta \cdot (\eta-4c_1), \\
 & n_{h_d} =  n_{h_{c1}} = n_{h_{c2}}=n_{h_u}= -2\eta \cdot (\eta-4c_1).
  \end{split}
\end{equation}
Given the base $B$ of elliptic fibration, $c_1,-t$ are determined purely by the property of $S$. Choosing $S$ such that $-\eta \cdot (\eta-4c_1)=3$, we
get three generations of SM fermions, plus six
pairs of doublet Higgses and six pairs of colored Higgses. Since we make use of the section $-c_1$ for $U(1)_Y$ we need $h^0( S, K_S ) > 0$. However it might imply the SM adjoint Higgses unless $S$ is torsion-free. 

We obtained the SM gauge group and three generations of quarks and
leptons {\em without} resorting to
an intermediate unification.
For the  vectorlike electroweak and colored Higgses,
we have the standard doublet-triplet splitting problem, also tightly
related to the $\mu$-problem.
We can further elaborate the model employing different factorization and/or
flux. Already this model has desirable symmetries to shed light on a
dynamical resolution.
Because of gauge invariance (\ref{lhe}) and the vanishing theorem, bare mass terms of F-theory
scale, close to the Planck scale, are forbidden. However below some
intermediate scale $M_I$ where the global symmetries and $U(1)_M$ are broken,
mass terms would be {\em dynamically} generated
\begin{equation}
 W = W_{\text{MSSM}}(\mu=0)+ m_{c} h_{c1} h_{c2}+ m_{h} h_u h_d.
\end{equation}
We expect some `standard solution' would generate the mass matrix
$m_{c}$ and $m_h$ at the scale $M_I$.
%Not being GUT, this model does not suffer problems like dimension six
%proton decay, and there is no need for the higher
%dimensional Higges for vacuum configuration.
We again emphasize the corresponding fields are distinguished by
$U(1)_M$ charges in Table
\ref{t:mcurve41}. We expect it would be
broken down to $\Z_2$ symmetry, becoming {\em matter parity}
\cite{KKK}. Then still
the terms (\ref{rviolating}) are forbidden in the low energy.
Also this structure can be related to dynamical supersymmetry
breaking \cite{Marsano:2009ym}.

The authors acknowledge the support
by the Grant-in-Aid for Scientific
Research No. 20$\cdot$08326 and 20540266 from the Ministry of
Education, Culture, Sports, Science and Technology of Japan (MEXT).
T.~K. is also supported in part by the Grant-in-Aid for the Global COE
Program ``The Next Generation of Physics, Spun from Universality and
Emergence'' from MEXT.


\begin{thebibliography}{99}

%\cite{Vafa:1996xn}
\bibitem{Vafa:1996xn}
  C.~Vafa,
  %``Evidence for F-Theory,''
  Nucl.\ Phys.\  B {\bf 469}, 403 (1996).
%  [arXiv:hep-th/9602022].
 %%CITATION = NUPHA,B469,403;%%

%\cite{Katz:1996xe}
\bibitem{KV}
  S.~H.~Katz and C.~Vafa,
  %``Matter from geometry,''
  Nucl.\ Phys.\  B {\bf 497}, 146 (1997).
%  [arXiv:hep-th/9606086].
  %%CITATION = NUPHA,B497,146;%%

%\cite{Donagi:2009ra}
\bibitem{Donagi:2009ra}
  R.~Donagi and M.~Wijnholt,
  %``Higgs Bundles and UV Completion in F-Theory,''
  arXiv:0904.1218 [hep-th].
  %%CITATION = ARXIV:0904.1218;%%


%\cite{Morrison:1996pp}
\bibitem{Morrison:1996pp}
  D.~R.~Morrison and C.~Vafa,
  %``Compactifications of F-Theory on Calabi--Yau Threefolds -- II,''
  Nucl.\ Phys.\  B {\bf 476}, 437 (1996);
%  [arXiv:hep-th/9603161].
  %%CITATION = NUPHA,B476,437;%%
%\cite{Morrison:1996na}
%\bibitem{Morrison:1996na}
%  D.~R.~Morrison and C.~Vafa,
  %``Compactifications of F-Theory on Calabi--Yau Threefolds -- I,''
  Nucl.\ Phys.\  B {\bf 473}, 74 (1996).
%  [arXiv:hep-th/9602114].
  %%CITATION = NUPHA,B473,74;%%

%\cite{Beasley:2008dc}
\bibitem{Beasley:2008dc}
  C.~Beasley, J.~J.~Heckman and C.~Vafa,
  %``GUTs and Exceptional Branes in F-theory - I,''
  JHEP {\bf 0901}, 058 (2009);
%  [arXiv:0802.3391 [hep-th]]
  %%CITATION = JHEPA,0901,058;%%
%\cite{Beasley:2008kw}
%\bibitem{Beasley:2008kw}
%  C.~Beasley, J.~J.~Heckman and C.~Vafa,
  %``GUTs and Exceptional Branes in F-theory - II: Experimental
  %Predictions,''
  JHEP {\bf 0901}, 059 (2009).
%  [arXiv:0806.0102 [hep-th]].
  %%CITATION = JHEPA,0901,059;%%

%\cite{Donagi:2008kj}
\bibitem{Donagi:2008kj}
  R.~Donagi and M.~Wijnholt,
  %``Breaking GUT Groups in F-Theory,''
  arXiv:0808.2223 [hep-th].
  %%CITATION = ARXIV:0808.2223;%%

%\cite{Heckman:2010bq}
\bibitem{Heckman:2010bq}
 See J.~J.~Heckman,
  %``Particle Physics Implications of F-theory,''
  arXiv:1001.0577 [hep-th] and references therein.
  %%CITATION = ARXIV:1001.0577;%%

%\cite{Choi:2009tp}
\bibitem{Choi:2009tp}
  K.~S.~Choi,
  %``Extended Gauge Symmetries in F-theory,''
  JHEP {\bf 1002}, 004 (2010).
  %%CITATION = JHEPA,1002,004;%%

\bibitem{KM}
S.~Katz and D.~R.~Morrison,
%``Gorenstein threefold singularities
%with small resolutions via invariant theory for Weyl groups'',
J. Algebraic Geom. {\bf 1} (1992), 449.

%\cite{Bershadsky:1996nh}
\bibitem{Bershadsky:1996nh}
  M.~Bershadsky et. al.
%, K.~A.~Intriligator, S.~Kachru, D.~R.~Morrison, V.~Sadov and C.~Vafa,
  %``Geometric singularities and enhanced gauge symmetries,''
  Nucl.\ Phys.\  B {\bf 481}, 215 (1996).
%  [arXiv:hep-th/9605200].
  %%CITATION = NUPHA,B481,215;%%

%\cite{Hayashi:2010zp}
\bibitem{Hayashi:2010zp}
  H.~Hayashi, T.~Kawano, Y.~Tsuchiya and T.~Watari,
  %``More on Dimension-4 Proton Decay Problem in F-theory -- Spectral Surface,
  %Discriminant Locus and Monodromy,''
  arXiv:1004.3870 [hep-th];
  %%CITATION = ARXIV:1004.3870;%%
%\cite{Grimm:2010ez}
%\bibitem{Grimm:2010ez}
  T.~W.~Grimm and T.~Weigand,
  %``On Abelian Gauge Symmetries and Proton Decay in Global F-theory GUTs,''
  arXiv:1006.0226 [hep-th].
  %%CITATION = ARXIV:1006.0226;%%

\bibitem{BMW}
In the heterotic language. See, e.g. 
%\cite{Blumenhagen:2006ux}
%\bibitem{Blumenhagen:2006ux}
  R.~Blumenhagen, S.~Moster and T.~Weigand,
  %``Heterotic GUT and standard model vacua from simply connected Calabi-Yau
  %manifolds,''
  Nucl.\ Phys.\  B {\bf 751}, 186 (2006).
%  [arXiv:hep-th/0603015].
  %%CITATION = NUPHA,B751,186;%%


%\cite{Friedman:1997yq}
\bibitem{Friedman:1997yq}
  R.~Friedman, J.~Morgan and E.~Witten,
  %``Vector bundles and F theory,''
  Commun.\ Math.\ Phys.\  {\bf 187}, 679 (1997).
%  [arXiv:hep-th/9701162].
  %%CITATION = CMPHA,187,679;%%

\bibitem{Abe:2009uz}
See also 
%\cite{Abe:2009uz}
  H.~Abe, K.~S.~Choi, T.~Kobayashi and H.~Ohki,
  %``Magnetic flux, Wilson line and orbifold,''
  Phys.\ Rev.\  D {\bf 80}, 126006 (2009);
%  [arXiv:0907.5274 [hep-th]].
  %%CITATION = PHRVA,D80,126006;%%
%\cite{Abe:2008sx} 
%%CITATION = NUPHA,B814,265;%%
%\cite{Choi:2009pv}
%\bibitem{Choi:2009pv}
  K.~S.~Choi et. al.,
%, T.~Kobayashi, R.~Maruyama, M.~Murata, Y.~Nakai, H.~Ohki and M.~Sakai,
  %``E6,7,8 Magnetized Extra Dimensional Models,''
  Eur.\ Phys.\ J.\  C {\bf 67}, 273 (2010).
%  [arXiv:0908.0395 [hep-ph]].
  %%CITATION = EPHJA,C67,273;%%


%\cite{Choi:2006hm}
\bibitem{Choi:2006hm}
  K.~S.~Choi,
  %``Unification in intersecting brane models,''
  Phys.\ Rev.\  D {\bf 74}, 066002 (2006);
%  [arXiv:hep-th/0603186].
  %%CITATION = PHRVA,D74,066002;%%
%\cite{Choi:2006th}
%\bibitem{Choi:2006th}
%  K.~S.~Choi,
  %``Intersecting brane world from type I compactification,''
  Int.\ J.\ Mod.\ Phys.\  A {\bf 22}, 3169 (2007);
%  [arXiv:hep-th/0610026].
  %%CITATION = IMPAE,A22,3169;%%
%\cite{Choi:2005pk}
%\bibitem{Choi:2005pk}
  K.~S.~Choi and J.~E.~Kim,
  %``Gauge unification via stable brane recombination,''
  JHEP {\bf 0511} (2005) 043.
%  [arXiv:hep-th/0508149].
  %%CITATION = JHEPA,0511,043;%%



%\cite{Hayashi:2009ge}
\bibitem{Hayashi:2009ge}
  H.~Hayashi, T.~Kawano, R.~Tatar and T.~Watari,
  %``Codimension-3 Singularities and Yukawa Couplings in F-theory,''
  Nucl.\ Phys.\  B {\bf 823}, 47 (2009).
%  [arXiv:0901.4941 [hep-th]].
  %%CITATION = NUPHA,B823,47;%%

%\cite{Marsano:2009gv}
\bibitem{Marsano:2009gv}
  J.~Marsano, N.~Saulina and S.~Schafer-Nameki,
  %``Monodromies, Fluxes, and Compact Three-Generation F-theory GUTs,''
  JHEP {\bf 0908}, 046 (2009).
%  [arXiv:0906.4672 [hep-th]].
  %%CITATION = JHEPA,0908,046;%%

%\cite{Choi:2010nf}
\bibitem{CK}
%\bibitem{Choi:2010nf}
  K.~S.~Choi,
  %``SU(3) x SU(2) x U(1) Vacua in F-Theory,''
  arXiv:1007.3843 [hep-th].
  %%CITATION = ARXIV:1007.3843;%%

%\cite{Blumenhagen:2009yv}
\bibitem{Blumenhagen:2009yv}
  R.~Blumenhagen, T.~W.~Grimm, B.~Jurke and T.~Weigand,
  %``Global F-theory GUTs,''
  Nucl.\ Phys.\  B {\bf 829}, 325 (2010).
%  [arXiv:0908.1784 [hep-th]].
  %%CITATION = NUPHA,B829,325;%%

%\cite{Hayashi:2008ba}
\bibitem{Hayashi:2008ba}
  H.~Hayashi, R.~Tatar, Y.~Toda, T.~Watari and M.~Yamazaki,
  %``New Aspects of Heterotic--F Theory Duality,''
  Nucl.\ Phys.\  B {\bf 806}, 224 (2009).
%  [arXiv:0805.1057 [hep-th]].
  %%CITATION = NUPHA,B806,224;%%

%\cite{Sadov:1996zm}
\bibitem{Sadov:1996zm}
  V.~Sadov,
  %``Generalized Green-Schwarz mechanism in F theory,''
  Phys.\ Lett.\  B {\bf 388}, 45 (1996).
%  [arXiv:hep-th/9606008].
  %%CITATION = PHLTA,B388,45;%%



\bibitem{KKK} See also
%\cite{Kim:2007mt}
%\bibitem{Kim:2007mt}
  J.~E.~Kim, J.~H.~Kim and B.~Kyae,
  %``Superstring standard model from Z(12-I) orbifold compactification with and
  %without exotics, and effective R-parity,''
  JHEP {\bf 0706}, 034 (2007);
%  [arXiv:hep-ph/0702278].
  %%CITATION = JHEPA,0706,034;%%
%\cite{Lebedev:2007hv}
%\bibitem{Lebedev:2007hv}
  O.~Lebedev et. al.
%, H.~P.~Nilles, S.~Raby, S.~Ramos-Sanchez, M.~Ratz, P.~K.~S.~Vaudrevange and A.~Wingerter,
  %``The Heterotic Road to the MSSM with R parity,''
  Phys.\ Rev.\  D {\bf 77} (2008) 046013.
%  [arXiv:0708.2691 [hep-th]].
  %%CITATION = PHRVA,D77,046013;%%

%\cite{Marsano:2009ym}
\bibitem{Marsano:2009ym}
  J.~Marsano, N.~Saulina and S.~Schafer-Nameki,
  %``F-theory Compactifications for Supersymmetric GUTs,''
  JHEP {\bf 0908} (2009) 030.
%  [arXiv:0904.3932 [hep-th]].
  %%CITATION = JHEPA,0908,030;%%







\end{thebibliography}
\end{document}